\documentclass{article}
\usepackage{spconf,amsmath,graphicx}

\title{Using joint training speaker encoder with consistency loss to achieve cross-lingual voice conversion and expressive voice conversion }
\name{Houjian Guo$^1$$^2$, Chaoran Liu$^3$, Carlos Toshinori Ishi$^1$$^3$, Hiroshi Ishiguro$^2$$^3$}
\address{$^1$Interactive Robot Research Team, Guardian Robot Project, RIKEN, Japan\\
  $^2$Graduate School of Engineering Science, Osaka University, Japan\\
  $^3$Advanced Telecommunications Research Institute International, Japan}
\begin{document}
%
\maketitle
\begin{abstract}
Voice conversion systems have made significant advancements in terms of naturalness and similarity in common voice conversion tasks. However, their performance in more complex tasks such as cross-lingual voice conversion and expressive voice conversion remains imperfect. In this study, we propose a novel approach that combines a jointly trained speaker encoder and content features extracted from the cross-lingual speech recognition model Whisper to achieve high-quality cross-lingual voice conversion. Additionally, we introduce a speaker consistency loss to the joint encoder, which improves the similarity between the converted speech and the reference speech. To further explore the capabilities of the joint speaker encoder, we use the Phonetic posteriorgram as the content feature, which enables the model to effectively reproduce both the speaker characteristics and the emotional aspects of the reference speech.  The code and pre-trained model are open-sourced \footnote{https://github.com/ConsistencyVC/ConsistencyVC-voive-conversion}.
\end{abstract}
\section{Introduction}
\label{sec:intro}

Voice conversion (VC) is a task that aims to modify a speaker's voice characteristics, such as speaker identity\cite{kaneko2018cyclegan},  emotion\cite{zhou2022emotional}, and accent\cite{wang2021accent} while preserving the linguistic content. In this study, we aim to address two challenges:

\begin{itemize}
\item Cross-lingual voice conversion (XVC), where the source speech and reference speech are in different languages\cite{zhou2019cross}.
\item Expressive voice conversion (EVC), which refers to the research conducted by Du et al.\cite{du2021expressive}, involves converting an input speech into a referenced speech in terms of both speaker identity and emotional style.
 \end{itemize}

Decoupling and reconstructing the information in speech is currently the most popular approach for high-quality voice conversion models\cite{van2022comparison}\cite{liu2021diffsvc}\cite{lee2022duration}. In detail, during training, content and speaker information is extracted from speech and then used for speech reconstruction. During inference, voice conversion is achieved by generating new speech using the content information from the source speech and the speaker information from the reference speech.

Phoneme posteriorgram (PPG) -based voice conversion, such as PPG-VC\cite{liu2021any}, is the classical voice conversion method based on this principle. However, in the past, limited automatic speech recognition (ASR) performance and insufficient speech synthesis model capability resulted in limited speech quality of the synthesized output\cite{sisman2020overview}. With the emergence of non-autoregressive (NAR) text-to-speech(TTS) models, such as VITS\cite{kim2021conditional}, FastSpeech2\cite{ren2020fastspeech}, DiffSinger\cite{liu2022diffsinger}, and the availability of large-scale pre-trained self-supervised learning (SSL) models, such as Hubert\cite{hsu2021hubert} and WavLM\cite{chen2022wavlm}, high-quality voice conversion models have become possible. The underlying principle of FreeVC\cite{li2023freevc}, ACE-VC\cite{hussain2023ace}, and the widely adopted open-source Singing Voice Conversion (SVC) model SO-VITS-SVC\footnote{https://github.com/svc-develop-team/so-vits-svc} is to leverage SSL models to extract content features from the original speech. Then, speaker ID or speaker classification models are employed to extract speaker-specific information from the speech. Finally,  both sets of information are used to reconstruct the speech by the TTS model. The quality of VC results depends heavily on the synthesis capabilities of the speech reconstruction model, especially when accurate and unambiguous content information is available.


However, there is potential to improve the method of extracting speaker features. Previous VC models, such as PPG-VC and Freevc, have typically used speaker encoders pre-trained on speaker classification tasks to acquire speaker embeddings that are then used to guide speech synthesis. It is important to note that the primary goal of training speaker encoders is not speech synthesis, but speaker recognition. Consequently, this approach may miss valuable information present in the reference speech, such as emotion. In addition, training speaker classification models requires large datasets.

Freevc-s\cite{li2023freevc}, Quickvc\cite{guo2023quickvc}, and NVC-Net\cite{nguyen2022nvc} use a jointly trained speaker encoder to ensure that the output of the speaker encoder contains only speaker-related information. This is achieved by implementing a bottleneck structure and carefully excluding speaker information from the content features. However, there is a lack of a more detailed loss function specifically designed to capture speaker-related features.

The concept of speaker consistency loss has been used in several studies, including YourTTS\cite{casanova2022yourtts} and CyclePPG-XVC\cite{zhou2021cross}, with the aim of improving the speaker similarity between model-generated speech and real speech. This is achieved by comparing the outputs of a speaker encoder that processes both the generated and the real speech. However, these studies used pre-trained speaker encoders that were specifically trained for speaker classification tasks. This method has certain limitations, such as the neglection of sentiment information, as discussed above. Furthermore, the speaker consistency loss used in these studies only updates the speech synthesis module and has no impact on the speaker encoder itself. Therefore, there is still room for improvement in the application of speaker consistency loss, especially for XVC and EVC tasks.

In this study, we introduce a novel VC model called ConsistencyVC to address issues related to speaker feature extraction. The main contributions of our research are summarised as follows:
\begin{itemize}
\item A new method for speaker feature extraction is proposed, where the speaker consistency loss is applied to the joint speaker encoder. Experimental results show that the inclusion of speaker consistency loss improves both speaker similarity and emotion similarity.

\item Implementation of XVC using Whisper: We use the intermediate features of the cross-lingual speech recognition model Whisper \cite{radford2022robust}, which can help to generate high-quality speech without foreign accents, to implement XVC.

\item Implementation of EVC with PPG as input: In order to imitate the emotional and speaker information in the reference speech, we implement EVC with PPG as input, which results in accurate conversion of both emotional and speaker characteristics.

 \end{itemize}

\label{sec:format}
\begin{figure*}[htbp]
    \centering
    \includegraphics[width=\textwidth]{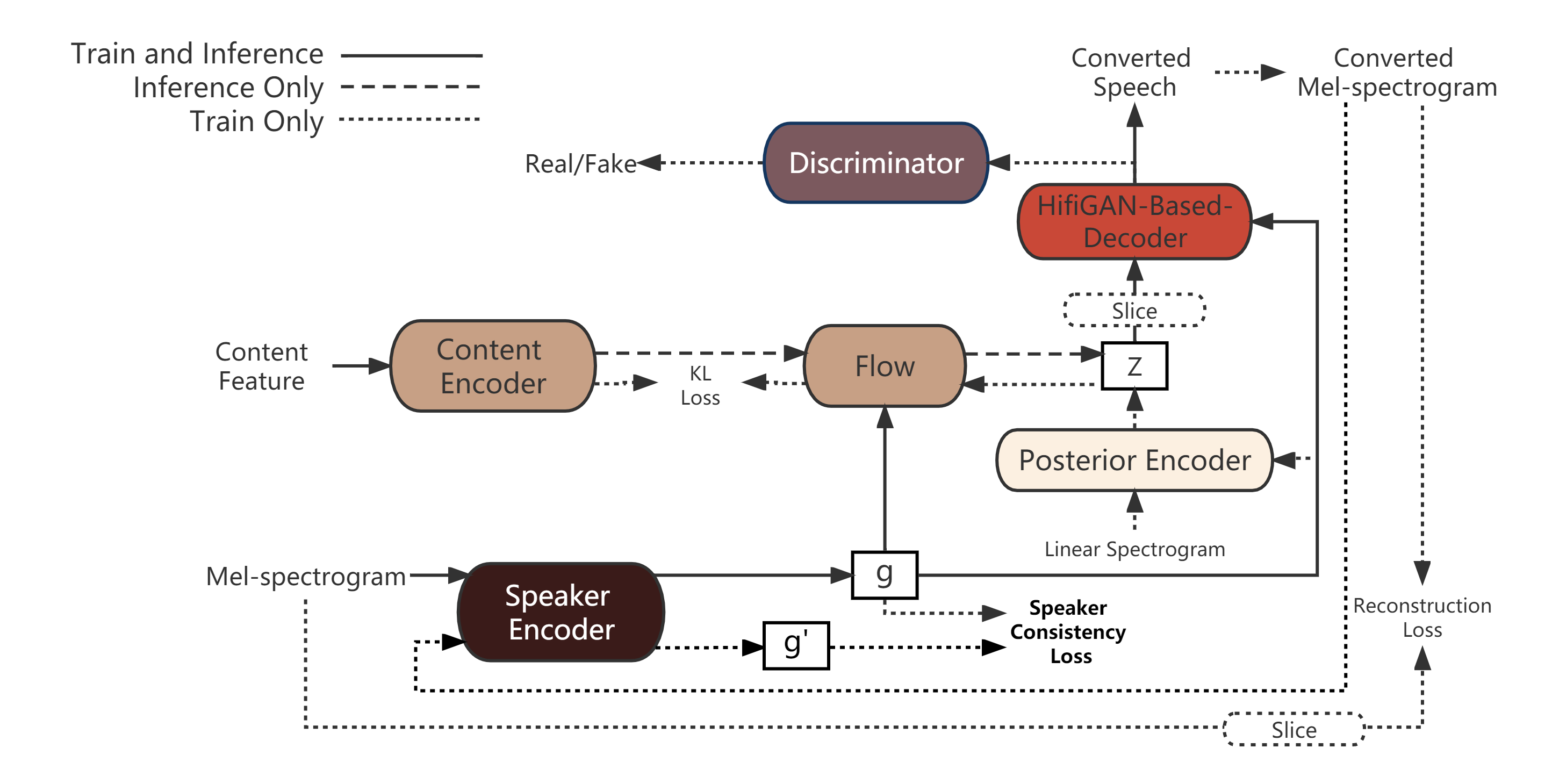}

    \caption{Overview network architecture of ConsistencyVC.  Here $g$ denotes speaker embedding, and $z$ denotes the latent variable. }
    \label{fig:speech_production}
\end{figure*}

\section{Method}
\subsection{Motivation}
In recent VC research, state-of-the-art VC systems have shown impressive performance, particularly in single-language (typically English) scenarios without emotion alteration. The speech samples that are generated show a remarkable degree of naturalness and similarity to human voices. We believe it is time to shift the focus of VC research from traditional VC to more complex applications. Therefore, we chose the XVC and EVC tasks to demonstrate the flexibility of our designed speaker information encoding approach, which can handle more tasks beyond traditional VC.

The proposed ConsistencyVC is inspired by FreeVC-s\cite{li2023freevc}, LoraSVC\footnote{https://github.com/PlayVoice/lora-svc}, VALL-e\cite{wang2023neural} and YourTTS\cite{casanova2022yourtts}. The model is based on the infrastructure of FreeVC-s due to its end-to-end structure, which enables high-quality VC. The bottleneck structure of the FreeVC-s's joint speaker encoder ensures that only speaker features are encoded, without including content information. In addition, the non-autoregressive design improves inference speed. However, unlike FreeVC-s, we take inspiration from LoraSVC and choose content features that perform better than the raw SSL model output for the XVC task. This eliminates the need for data augmentation, which consumes a lot of storage space. VALL-e is a zero-shot speech synthesis model that uses 3-second segments of target speech as reference speech input during training. The model can mimic the speaker's features and even emotional features from any 3-second reference speech segment and synthesize high-quality speech. This slicing concept inspired us to assume that a sub-segment of speech should contain similar speaker and emotional features as the whole speech. Building on this assumption, and inspired by YourTTS, we apply speaker consistency loss to the training of the joint speaker encoder VC model. This application of speaker consistency loss improves speaker similarity and emotion similarity.

\subsection{Model architecture}
As shown in Figure 1, the main structure of the ConsistencyVC model follows the VITS speech synthesis model. However, the text encoder and duration predictor are replaced by a content encoder similar to the posterior encoder in VITS. The content encoder uses WaveNet residual blocks. Instead of text, the content encoder takes content features from a pre-trained ASR model as input. Speaker information is encoded from Mel-spectrogram using a jointly trained speaker encoder. In the experiment part, the XVC and EVC tasks use different datasets and different types of content features to train the VC model. However, the structure of the VC model remains the same.
In the following, the content feature selection and the structure of the speaker encoder are explained in detail.
\subsubsection{Content feature}
Whisper is an ASR model that has achieved remarkable results in cross-lingual speech recognition tasks. Its model architecture is based on an encoder-decoder transformer. We choose the output of the transformer encoder blocks, called Whisper Encoder's Output (WEO), as the content feature. Compared to the content features in previous XVC studies, the WEO provides more accurate and comprehensive information, including the accent of the source speech, which is crucial for achieving foreign accent-free XVC. Therefore, we choose WEO as the content feature for the XVC task. 

For the EVC task, we choose to use the PPG obtained from the wav2vec model trained on the phoneme recognition task\footnote{https://huggingface.co/speech31/wav2vec2-large-english-TIMIT-phoneme\_v3}. This choice is based on the fact that different emotions within the same utterance have different prosody. PPG provides clear content information and contains less prosodic information than WEO, so the prosody of the reconstructed speech relies entirely on the output of the speaker encoder. This makes PPG more suitable for EVC.

Conversely, WEO is more suitable than PPG for XVC. This is because the same pronunciation may have different prosody or accent in different languages. In the XVC task, the prosody of the reconstructed speech should depend on the source speech, which comes from a native speaker.
\subsubsection{Speaker Encoder}
The coded speaker embedding is generated by the speaker encoder using the Mel-spectrogram as input. The speaker encoder is trained together with the rest of the model. It consists of one block of 3-layer LSTM module and a fully connected layer, similar to FreeVC-s. We feed the Mel-spectrogram derived from the speech into the LSTM layer of the speaker encoder. The final hidden state of the LSTM is passed to the fully connected layers, allowing the transformation of variable-length Mel-spectrogram inputs into fixed-size embeddings, which achieves a bottleneck structure. The output of the content encoder is assumed to be speaker independent. To synthesize speech, the model replaces the missing speaker information by using the input from the speaker encoder.

\subsection{Training strategy}

Following the training strategy of VITS, the ConsistencyVC model incorporates VAE and adversarial training during the training process.
For the generator part, the loss can be expressed as:
\begin{equation}
  {\cal L}_{v a e}={\cal L}_{r e c o n}+{\cal L}_{k l}+{\cal L}_{a d v}(G)+{\cal L}_{f m}(G)+{\cal L}_{S C L},
  \label{eq0}
\end{equation}
where the ${\cal L}_{r e c o n}$ is the reconstruction loss, the ${\cal L}_{k l}$ is the KL loss, ${\cal L}_{a d v}(G)$ is the adversarial loss, and ${\cal L}_{f m}(G)$ is the feature matching loss. These losses are similar to the VITS, so the specific details of these losses are not reiterated here. Instead, let's focus on explaining the speaker consistency loss ${\cal L}_{S C L}$.

In the implementation of VITS, researchers adopt windowed generator training\cite{ren2019fastspeech}, a technique that generates only a part of the raw waveforms during training to reduce computational requirements. They randomly extract segments of latent representations $z$ to feed into the HiFi-GAN-based decoder, and corresponding audio segments are extracted from the ground truth raw waveforms as training targets. This leads to the model's output speech's length during training being a small part of  the input speech's length.

In FreeVC-S, for the VC task, the content information input is also segmented, limiting the maximum size of the content features. We assume that the speech segment corresponding to the input content features and the speech segment output by the model should contain the same emotion and speaker features. Based on this assumption, we can design the speaker consistency loss for the jointly trained speaker encoder.

Formally, let $\phi(\cdot)$ be the function of the speaker encoder that outputs the speaker embedding of the reference,  The speaker consistency loss is defined as the L1 distance between the speaker embeddings of the ground truth speech segmentation and the generated speech segmentation:

\begin{equation}
{\cal L}_{S C L} = \Vert \phi(t) - \phi(h) \Vert_{1} .
\label{eq1}
\end{equation}

where $t$ and $h$ represent, respectively, the ground truth speech segmentation and the generated speech segmentation. Similar to YourTTS and other research, we do not introduce the speaker consistency loss at the beginning of the training step but rather after the model has learned the basic speech synthesis capability. 

As far as our knowledge goes, we are the first to introduce the consistency loss to the joint-trained speaker encoder.
\section{Experiment}
\label{sec:pagestyle}
\begin{table*}[htbp]
\centering
\caption{Naturalness MOS of XVC}
\begin{tabular}{c|cc|cc}
\hline
                        & \multicolumn{2}{c|}{Seen Speaker}                       & \multicolumn{2}{c}{Unseen Speaker}                      \\ \hline
                        & \multicolumn{1}{c|}{Same Language} & Different Language & \multicolumn{1}{c|}{Same Language} & Different Language \\ \hline
BNE-PPG-VC              & \multicolumn{1}{c|}{3.10$\pm$0.12}             & 2.67$\pm$0.10                  & \multicolumn{1}{c|}{3.26$\pm$0.13}             & 2.81$\pm$0.13                  \\ 
ConsistencyXVC-w/o loss       & \multicolumn{1}{c|}{3.97$\pm$0.10}             & 3.92$\pm$0.08               & \multicolumn{1}{c|}{3.95$\pm$0.13}             & 3.99$\pm$0.10                  \\ 
ConsistencyXVC& \multicolumn{1}{c|}{\textbf{4.09$\pm$0.10}}             & \textbf{4.08$\pm$0.09}                  & \multicolumn{1}{c|}{\textbf{4.31$\pm$0.10}}             & \textbf{4.16$\pm$0.10}                  \\ \hline
Ground Truth  &\multicolumn{4}{c}{4.27$\pm$0.10}           \\ \hline
\end{tabular}
\end{table*}

\begin{table*}[htbp]
\centering
\caption{Similarity MOS of XVC}
\begin{tabular}{c|cc|cc}
\hline
                        & \multicolumn{2}{c|}{Seen Speaker}                       & \multicolumn{2}{c}{Unseen Speaker}                      \\ \hline
                        & \multicolumn{1}{c|}{Same Language} & Different Language & \multicolumn{1}{c|}{Same Language} & Different Language \\ \hline
BNE-PPG-VC              & \multicolumn{1}{c|}{2.73$\pm$0.25}             & 1.97$\pm$0.13                  & \multicolumn{1}{c|}{2.52$\pm$0.23}             & 2.39 $\pm$0.22                 \\ 
ConsistencyXVC-w/o loss       & \multicolumn{1}{c|}{3.54$\pm$0.20}             & 3.36$\pm$0.15                  & \multicolumn{1}{c|}{\textbf{3.13$\pm$0.21}}             & \textbf{2.77$\pm$0.22}                  \\ 
ConsistencyXVC& \multicolumn{1}{c|}{\textbf{4.12$\pm$0.18}}             & \textbf{3.64$\pm$0.14}                  & \multicolumn{1}{c|}{3.06$\pm$0.21}             & 2.70$\pm$0.21                  \\ \hline
\end{tabular}
\end{table*}
\subsection{Cross-lingual Voice Conversion}

\subsubsection{Dataset}
In the XVC experiment, we used several datasets, including Aishell-3\cite{shi2020aishell}, LibriTTS-100\cite{zen2019libritts}, JVS\cite{takamichi2019jvs}, ESD\cite{zhou2022emotional}, VCTK\cite{veaux2017cstr}, Aishell-1\cite{bu2017aishell}, and JECS\cite{nakayama2018japanese}. Of these, LibriTTS-100, ESD, Aishell3, and JVC contain speech samples in English, Chinese, and Japanese; these datasets were used to train the XVC model. The VCTK, Aishell-1, and JECS datasets were used to provide unseen speaker samples in English, Chinese, and Japanese respectively. These unseen speaker samples were used to evaluate the ability of the model to imitate speakers not present in the training set.

\subsubsection{Experimental setup}
For our experiments, we used a sampling rate of 16,000 Hz. The utterances of each speaker were randomly split into training and test sets in a 9:1 ratio. The model is based on FreeVC-s, but there are some parameters designed differently.

The most significant difference is that for the XVC task we choose WEO as the content information, the hop size of WEO is 320. In terms of other inputs to the model, both linear spectrogram and 80-band Mel-spectrogram are computed using Short-Time Fourier Transform (STFT) with FFT size, window size, and hop size being set to 1024, 1024 and 320 respectively.  The upsampling scale of the four residual blocks in the HiFi-GAN-based decoder is factorized as 320 = 10 × 8 × 2 × 2, which means that the upsampling scales for the four blocks are [10, 8, 2, 2]. To avoid potential checkerboard artifacts caused by the "ConvTranspose1d" upsampling layer\cite{odena2016deconvolution}, kernel sizes of [20, 16, 4, 4] are used. The AdamW optimizer is used with the same weight decay and learning rate as in FreeVC-s.

In our experiments, we compare two versions of ConsistencyVC: ConsistencyXVC and ConsistencyXVC-w/o loss. Both models are trained on a single NVIDIA 3090 GPU for up to 300k steps using fp16 training. For the first 100k steps, the batch size is 108, and the utterance lengths used for training range from 24,000 to 96,000 samples, corresponding to 1.5 to 6 seconds. The latent variables z input to the HiFi-GAN-based decoder are sliced into 28 segments, resulting in a speech length of 28 × 320 = 8,960 samples. However, for ConsistencyXVC, starting from the 100k step mark, an additional training phase with speaker consistency loss is introduced for the next 200k steps. In this phase, the latent variables z are sliced into 75 segments, resulting in a speech length of 75 × 320 = 24,000 samples. Due to the larger size of the latent variables z, the batch size is reduced to 42. ConsistencyXVC-w/o loss, on the other hand, continues training with the same parameters without introducing speaker consistency loss up to 300k steps.

We also chose BNE-PPG-VC as a baseline, which uses F0, PPG and speaker embeddings as inputs to a seq2seq model for speech reconstruction. Since the model has access to the F0 information of the source speech, it is also able to perform XVC without foreign accents. We trained BNE-PPG-VC on the same tri-lingual dataset as ConsistencyXVC.

\subsubsection{Subjective evaluation}
In the subjective experiments, we adopted the Mean Opinion Score (MOS) as the subjective metric to compute the naturalness and similarity scores of the converted utterances. We invited 47 native English speakers from Amazon Mechanical Turk \footnote{https://requester.mturk.com} to evaluate the speech. All speech's source speech is English but reference speech is English, Chinese or Japanese.
Each subject was required to evaluate the naturalness of 6 original utterances from the dataset and 60 converted utterances. Additionally, they were asked to evaluate the similarity of 32 converted utterances to the utterances of the target speakers. Some audio samples are available on the demo page\footnote{https://consistencyvc.github.io/ConsistencyVC-demo-page/}.

The experimental results for naturalness in Table 1 show the fact that the reference voice is not English does not affect the naturalness of English. This indicates that the model is successful in the XVC task. Meanwhile, the speaker similarity experiments show that the introduction of speaker consistency loss improves the similarity between the speakers of the generated speech and the speakers that appeared in the model's training set. However, for speakers not seen in the training set, speaker consistency loss did not serve to improve speaker similarity.

\subsection{Expressive Voice Conversion}
\begin{table}[]
\centering
\caption{Naturalness MOS of EVC, seen means seen speaker, unseen means unseen speaker}
\begin{tabular}{c|c|c}
\hline
                        & Seen & Unseen \\ \hline
ConsistencyEVC          & 3.52$\pm$0.08            & 3.63$\pm$0.14              \\ 
ConsistencyEVC-w/o loss & 3.26$\pm$0.08            & 3.52$\pm$0.14              \\ 
ConsistencyEVC-whisper  & \textbf{3.95$\pm$0.07}            & \textbf{3.92$\pm$0.11}              \\ \hline
Ground Truth  &\multicolumn{2}{c}{4.20$\pm$0.08}           \\ \hline
\end{tabular}
\end{table}

\subsubsection{Dataset}

For the EVC task, we conducted the experiments using English datasets. We selected the English data from the ESD dataset\cite{zhou2022emotional} and the VCTK dataset\cite{veaux2017cstr} to train the models. Additionally, we selected samples from the Emov-db dataset\cite{adigwe2018emotional} as the reference speech to consider the model's ability to imitate emotional speech from speakers that were not present in the training set.
\subsubsection{Experimental setup}
In the EVC task, we compared different variations of the ConsistencyVC model, including the presence or absence of speaker consistency loss, and using different types of content features, including PPG and WEO.

The variations of the model that were compared are as follows:

1. ConsistencyEVC: This model uses PPG as the content feature input. It is trained without speaker consistency loss for the first 100k steps with a batch size of 108. Then, it continues training with speaker consistency loss for the next 200k steps with a reduced batch size of 42.

2. ConsistencyEVC-w/o loss: This model also uses PPG as the content feature input. It is trained without speaker consistency loss for the entire duration of 300k steps with a batch size of 108.

3. ConsistencyEVC-whisper: This model uses WEO as the content feature input. Similar to ConsistencyEVC, it is trained without speaker consistency loss for the first 100k steps with a batch size of 108. Then, it continues training with speaker consistency loss for the next 200k steps with a reduced batch size of 42.

Apart from the dataset and content features, all other training parameters remain the same as those used in the XVC task.

\subsubsection{Subjective evaluation}
In the subjective experiments, following Du et al.\cite{du2021disentanglement}, we used the Mean Opinion Score (MOS) to calculate the naturalness of the speech and the ABX preference test to compare the results of the different methods in terms of style similarity. We invited 45 subjects from Amazon Mechanical Turk to participate in the experiment. Six original utterances from the dataset and 60 converted utterances were evaluated by each subject for their naturalness, and there are 36 sets of preference tests. In each test, the subjects were asked to choose which utterance is more similar to the reference utterance, in terms of both speaker and emotion.

Similar to the XVC task, the use of speaker consistency loss improved the model's ability to imitate the reference speech of the seen speakers. Furthermore, the experimental results showed that WEO as a content feature outperformed PPG in terms of naturalness as measured by MOS. WEO contains more information suitable for reconstruction, resulting in more natural synthesized speech.

However, in the ABX preference test, ConsistencyEVC-whisper performed worse than ConsistencyEVC. This is because WEO contains additional information beyond content, such as intonation, which is helpful for XVC tasks to eliminate accents when foreign speakers speak in a different language. For EVC tasks, however, we want to retain only the content features of the source speech and let the reference speech determine the intonation of the converted speech.

\begin{figure}[]
    \centering
    \includegraphics[width=0.4\textwidth]{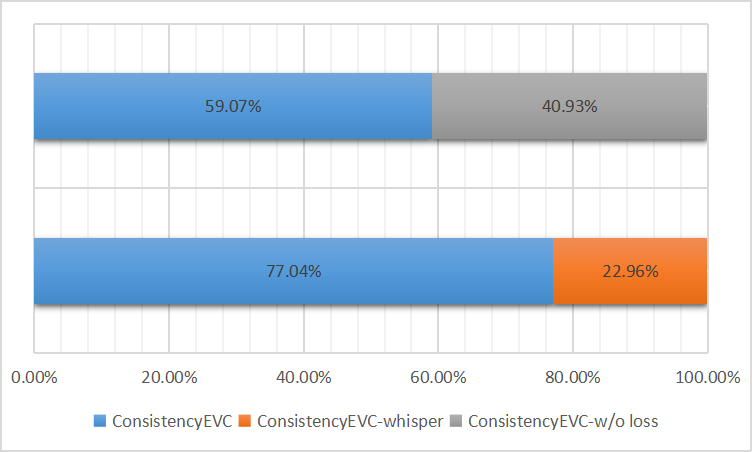}
   
    \caption{The ABX preference results for seen speakers of EVC.}
\end{figure}

\begin{figure}[]
    \centering
    \includegraphics[width=0.4\textwidth]{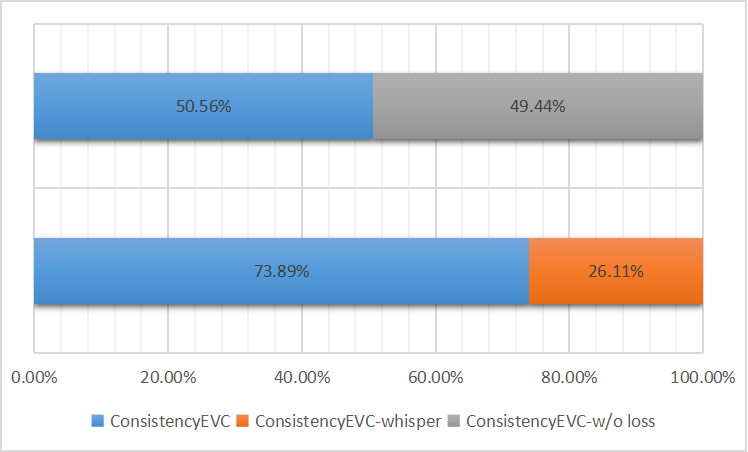}
   
    \caption{The ABX preference results for unseen speakers of EVC.}
\end{figure}

\section{Conclusion, Limitations, and Future Work}
\label{sec:typestyle}

In this study, we built on FreeVC-s to implement cross-lingual voice conversion and expressive voice conversion using the cross-lingual speech recognition model Whisper and the wav2vec-based phoneme recognition model, respectively. To improve the imitation of reference speech, we introduced the speaker consistency loss to the joint speaker encoder. Experimental results showed that this loss contributed to improvements in both speaker and emotion features. However, there are still some limitations to our research:

Speaker similarity decreases when the reference speech belongs to an unseen speaker in the dataset. Training the model with a more diverse set of speakers could potentially improve its ability to imitate unseen speakers. For example, training with the LibriTTS-R dataset\cite{koizumi2023libritts}, which consists of 585 hours of speech data from 2,456 speakers, could allow for better zero-shot and higher-quality voice conversion. 

The content features for XVC and EVC tasks are inconsistent. The choice of content features is flexible. If the focus is on speech quality and maintaining a similar pitch in the converted speech, then Whisper can be chosen. However, if the focus is on emotional expression in speech, the use of PPG as content features can ensure that the VC model generates speech with the same style as the reference speech. It is likely that the two tasks are not mutually exclusive, as XVC requires the preservation of intonation, which needs to be modified in the EVC tasks. In future research, it would be worthwhile to explore approaches to decouple emotion and speaker information in speech. 


\bibliographystyle{IEEEbib}
\bibliography{strings,refs}

\end{document}